\documentclass[letterpaper,twocolumn,english,showpacs,prl,aps]{revtex4}
\usepackage[T1]{fontenc}
\usepackage[latin1]{inputenc}
\usepackage{graphicx}
\usepackage{amssymb}

\makeatletter

\large  

\input epsf

\usepackage{babel}
\makeatother
\begin{document}

\title{Quantum dissociation of an edge of a Luttinger liquid}

\author{Eugene B. Kolomeisky and Michael Timmins}

\affiliation{Department of Physics, University of Virginia, P. O. Box 400714,
Charlottesville, Virginia 22904-4714}

\begin{abstract}
In a Luttinger liquid phase of one-dimensional molecular matter
the strength of zero-point 
motion can be characterized by dimensionless De Boer's number
quantifying the interplay of quantum fluctuations and two-body
interactions.  Selecting the latter in the Morse form we show that
dissociation of the Luttinger liquid is a process initiated at the
system edge.  The latter becomes unstable against quantum fluctuations at a
value of De Boer's number which is smaller than that of the bulk
instability which parallels the classical phenomenon of surface melting.

\end{abstract}


\maketitle

A classical three-dimensional solid melts through a first-order
transition at a temperature when the free energies of
the solid and liquid phases coincide.  At sufficiently low
temperatures quantum effects dominate and a quantum
solid can melt due to zero-point motion \cite{Anderson}.

The most curious feature of classical melting is the difficulty in
overheating the solid while supercooling the liquid is easy.  The
latter is expected for the first-order transition while the former is
explained by the phenomenon of surface melting: often, as the bulk transition
is approached, the melting begins at the free surface of
a solid.  The surface melting is well-documented experimentally, and
phenomenologically it can be viewed as a wetting of the
solid by its own melt \cite{SM}.    

A well-understood example of surface melting of a quantum solid is
that of the edge melting of the two-dimensional Wigner crystal in a
strong magnetic field \cite{Fertig}   

The goal of this note is to point out that strictly one-dimensional matter
with a free edge can also exhibit an analog of surface melting.
Fundamentally this happens because the edge represents a zero-dimensional system subject
to stronger quantum fluctuations than the one-dimensional bulk.
Due to broken translational symmetry, zero-point
motion modifies the cohesive properties of the edge differently from those
of the bulk.           

Experimentally one-dimensional matter can be realized in carbon
nanotube bundles \cite{Iijima}.  The latter can play a role of 
one-dimensional hosts for foreign particles that can find themselves
bound in the interstitial channels or inside the tubes
\cite{nanotubes2}.  Additionally one-dimensional atomic chains can be
constructed on selected templates with the help of scanning tunneling
microscopy, or via self-assembly of the deposited material of the
chain \cite{chain}.   

Consider a many-body system of identical particles of mass
$m$ with pairwise interaction $V(h)$ corresponding
to the molecular matter \cite{Ashcroft&Mermin}:  at large interparticle
separation $h$ the interaction is dominated by weak rapidly decaying
van der Waals attraction, while at short distances there is a
strong overlap repulsion . As a result, the pair potential $V(h)$ has
an asymmetric minimum at some intermediate $h$.  Assume that the pair 
potential is of the form 
\begin{equation}
\label{potential}
V(h)=\epsilon U(h/l - Q_{0}),
\end{equation}
where $\epsilon$ is the energy scale of the potential, $l$ is the
potential range, $Q_{0}$ is a dimensionless parameter, and $U(y)$ is a function
common to a family of substances.  With this choice the quantum
theorem of the corresponding states \cite{deBoer} holds stating
that every property measured in appropriate dimensionless units is
only determined by the function $U(y)$, particle
statistics and De Boer's number 
\begin{equation}
\label{lambda_{0}}
\lambda_{0}={\frac{\hbar}{\pi l(2m\epsilon)^{1/2}}},
\end{equation}
measuring the intensity of zero-point motion.

The possibility of several bulk phases in the system translates into the
corresponding number of the branches of the energy as a function of 
$\lambda_{0}$;  the lowest of them singles
out the ground state of the system. When two energy curves
cross, the ground state changes
via a first-order phase transition. For sufficiently large
$\lambda_{0}$ and zero pressure the ground state must correspond to
individual particles infinitely far apart from each other. This is a
monoatomic gas which will be chosen as the zero reference point for the energy.

In what follows we select the pair interaction potential in the Morse
form \cite{Morse}: 
\begin{equation}
\label{Morse}
V(h)  =  \epsilon(e^{-2(h/l - Q_{0})} - 2e^{-(h/l -Q_{0})}),
\end{equation}
where $\epsilon$ is the depth of the potential well and $Q_{0}$ is the location
of the minimum of (\ref{Morse}) measured in units of the potential 
range $l$.  Similar to the applications of
the Lennard-Jones potential to laboratory molecular systems 
\cite{Ashcroft&Mermin}, the only reason behind this choice is the
possibility of analytic progress.  Morse parameters for a series of
molecular substances and corresponding De Boer's numbers (\ref{lambda_{0}})
were computed in Ref.\cite{KQT}.  Hereafter the energy
and length scales will be measured in units of $\epsilon$ and $l$, 
respectively.  As appropriate for molecular substances, we 
restrict ourselves to nearest-neighbor interactions.  

In the classical limit, $\lambda_{0} = 0$, the ground state of the
system is a crystal; its quantum counterpart for sufficiently small 
$\lambda_{0}$ is a Luttinger liquid \cite{Luttinger} whose properties have been
computed in Ref.\cite{KQT} as follows:

The length of any bulk bond $h$ as a function of imaginary time $\tau$ 
is viewed as a quantum-mechanical degree of freedom subject to the
external potential $V(h)$.  This bond joins together two  
half-infinite segments representing the rest of the system, the ``bath''.  After the bath is approximated
by a harmonic liquid, the latter can be integrated out away from the 
anharmonic bond leading to a problem of the Caldeira-Leggett type 
\cite{Caldeira&Leggett}.  The latter has been analyzed by a
combination of variational and renormalization-group techniques, and
it has been demonstrated that the approximation is a controlled way of 
dealing with the interplay of zero-point motion and anharmonicity of
the two-body interaction \cite{KQT}.   Similar consideration applied
to the edge bond of a half-infinite Luttinger liquid leads to the
Euclidian action of the form
\begin{equation}
\label{wactionedge}
S_{edge}={\frac{\rho c}{8}}\int\limits
_{|\omega| < \omega_{D}}{\frac{d\omega}{2\pi}}|\omega||h(\omega)|^{2}+\int
d\tau V(h),
\end{equation}
where $\rho$ and $c$ are the mass density and sound velocity,
respectively, and $h(\omega)$ is the Fourier
transform of the bond-length field; the frequency cutoff is given by
the Debye frequency $\omega_{D}$.

The $\rho c|\omega|$ form of the kinetic energy term of the
action (\ref{wactionedge}) can be understood by noticing that if the
bond length oscillates with frequency 
$\omega$, then during one oscillation period $2\pi/|\omega|$ this 
disturbance propagates in the bulk a distance of order
$c/|\omega|$. Thus the usual kinetic energy
density, proportional to $\rho\omega^{2}$ should be multiplied by the
size of the region $c/|\omega|$ affected by the motion.

The calculation of the properties of the edge of a Luttinger liquid
proceeds through the application to the action (\ref{wactionedge}) of  
Feynman's variational principle \cite{Feynman} which states that for
any trial action $S_{0}$ with associated ground-state energy $E_{0}$,
the system's true ground-state energy is bounded above by 
$E_{0}+(T/\hbar)<S-S_{0}>_{0}$ where the zero-temperature limit
$T=0$ is taken at the end and $<>_{0}$ denotes an expectation value
computed with $S_{0}$.

Similar to the bulk problem \cite{KQT} the trial action is selected in
the Gaussian form 
\begin{equation}
\label{trialaction}
S_{0}={\frac{\rho c}{8}}\left (\int\limits
_{|\omega| \le \omega_{D}}{\frac{d\omega}{2\pi}}|\omega||h(\omega)|^{2}+
\gamma \omega_{D}\int
d\tau (h-Ql)^{2}\right ),
\end{equation}
where dimensionless variational parameters $Q$ and $\gamma$ have a
meaning of the bond length and its stiffness, respectively.  Then the
root-mean-square (rms) fluctuation of the bond length can be computed
as
\begin{equation}
\label{rmsedge}
<f^{2}>_{0~edge}^{1/2} = 2\lambda^{1/2}\ln^{1/2}(1+\gamma^{-1}),
\end{equation} 
where  
\begin{equation}
\label{lambda}
\lambda={\frac{\hbar}{\pi\rho cl^{2}}}
\end{equation}
quantifies the strength of zero-point motion in the Luttinger liquid.
The binding energy of the edge particle $E_{edge}$ is approximated by
$E_{0}+(T/\hbar)<S-S_{0}>_{0}$, i. e. by its upper bound: 
\begin{eqnarray}
\label{eofgammaqedge}
E_{edge}(\gamma,Q) & = & (\pi \lambda_{0}^{2}/\lambda)\ln(1+\gamma)-2e^{Q_{0}-Q}(1+\gamma^{-1})^{2\lambda}\nonumber \\
 & + & e^{2(Q_{0}-Q)}(1+\gamma^{-1})^{8\lambda}
\end{eqnarray}
Minimizing $E_{edge}$ with respect to $Q$ we arrive at the expression
for the quantum expansion of the edge bond
\begin{equation}
\label{qexpansionedge}
Q_{edge} - Q_{0} = 6\lambda\ln(1+\gamma^{-1})
\end{equation}
Substituting this back into (\ref{eofgammaqedge}),
$E_{edge}$ can be written as  
\begin{equation}
\label{eofgammaedge}
E_{edge}(\gamma)=(\pi\lambda_{0}^{2}/\lambda)\ln(1+\gamma)-(1+\gamma^{-1})^{-4\lambda}
\end{equation}
Minimizing Eq.(\ref{eofgammaedge}) with respect to $\gamma$, and
substituting the outcome back into (\ref{eofgammaedge}) we find 
\begin{equation}
\label{gammaedge1}
\gamma = (4\lambda^{2}/\pi \lambda_{0}^{2})(1+\gamma^{-1})^{-4\lambda}
\end{equation}
and
\begin{equation}
\label{eedge1}
E_{edge} = (\pi\lambda_{0}^{2}/4\lambda^{2})\left (4\lambda\ln(1 +
\gamma) - \gamma\right )
\end{equation}
respectively.  The results (\ref{rmsedge}), (\ref{qexpansionedge}),
(\ref{gammaedge1}), and (\ref{eedge1}) should be compared with their bulk
counterparts \cite{KQT}:   
\begin{equation}
\label{rmsbulk}
<f^{2}>^{1/2}_{0~bulk}=(2\lambda)^{1/2}\ln^{1/2}(1+\pi/2),
\end{equation}
\begin{equation}
\label{qexpansionbulk}
Q_{bulk} - Q_{0} = 3\lambda\ln(1+\pi/2),
\end{equation}
\begin{equation}
\label{lambda_0versuslambda}
\lambda_{0}=\lambda (1+\pi/2)^{-\lambda}
\end{equation}
\begin{equation}
\label{energybulk}
E_{bulk} = (1+\pi/2)^{-2\lambda}\left (\pi\lambda\ln(1+2/\pi)-1 \right )
\end{equation}
Substituting Eq.(\ref{lambda_0versuslambda}) back in
Eqs.(\ref{gammaedge1}) and (\ref{eedge1}) brings them into a form
convenient for analysis
\begin{equation}
\label{gammaedge2}
\gamma = (4/\pi)(1 + \pi/2)^{2\lambda}(1+\gamma^{-1})^{-4\lambda}
\end{equation}
\begin{equation}   
\label{eedge2}
E_{edge} = (\pi/4)(1 + \pi/2)^{-2\lambda}\left (4\lambda\ln(1 +
\gamma) - \gamma\right )
\end{equation}
The properties of the edge as a function of the quantum parameter
$\lambda$ (\ref{lambda}) can be computed by finding a solution
$\gamma(\lambda)$ to Eq.(\ref{gammaedge2}) minimizing the energy
(\ref{eedge2}) and substituting the outcome in the expressions for the
rms fluctuation (\ref{rmsedge}) and quantum expansion
(\ref{qexpansionedge}); the dependence on De Boer's number
(\ref{lambda_{0}}) follows from Eq.(\ref{lambda_0versuslambda}).

In the classical limit, $\lambda \rightarrow 0$, the only solution to
(\ref{gammaedge2}) is $\gamma = 4/\pi$ with the energy (\ref{eedge2})
$E_{edge} = -1$ as expected.  As the degree of zero-point motion
intensifies ($\lambda$ increases), the bond stiffness $\gamma$
decreases and the energy $E_{edge}$ increases.  For finite $\lambda$
Eq.(\ref{gammaedge2}) may have more than one solution. One of them is
always $\gamma =0$ corresponding to the delocalized edge particle.  For
large $\lambda$ this solution must correspond to the lowest (zero)
energy (\ref{eedge2}).

For $\gamma \ll 1$ the right-hand-side of Eq.(\ref{gammaedge2}) behaves
as $\gamma^{4\lambda}$ while for $\gamma \rightarrow \infty$ it
approaches a $\gamma$-independent limit, thus implying that
(\ref{gammaedge2}) cannot have more than three solutions and that
$\lambda = 1/4$ plays a special role.

For $0 < \lambda \le 1/4$ Eq.(\ref{gammaedge2}) has two solutions and
the larger of them (whose $\lambda = 0$ limit is $\gamma = 4/\pi$)
corresponds to the lowest energy (\ref{eedge2}).  For $\lambda = 1/4$
the explicit solution to (\ref{gammaedge2}) is $\gamma = (4/\pi)(1 +
\pi/2)^{1/2} - 1 \simeq 1.0415$

As $\lambda$ increases beyond $1/4$, Eq.(\ref{gammaedge2}) acquires a
third root whose $\lambda \rightarrow 1/4+0$ limit is $\gamma =
(\pi/4(1 + \pi/2)^{1/2})^{1/(4\lambda - 1)} \rightarrow 0$.  However
this solution leads to a larger energy (\ref{eedge2}) than even the delocalized
solution $\gamma = 0$.  The lowest energy (bound) state continues to
be described by the largest solution to (\ref{gammaedge2}).

As $\lambda$ continues to increase, the finite solutions to
(\ref{gammaedge2}) approach each other and at some $\lambda$ they
coalesce.  This is a critical phenomenon corresponding to the limit of
stability of the bound edge.  At that point the slopes of the right-
and left-hand-sides of Eq.(\ref{gammaedge2}) coincide which leads to
the limiting values $\gamma \simeq 0.4920$ and $\lambda \simeq 0.3730$
satisfying the relationship $\gamma = 4\lambda - 1$.  At larger values
of $\lambda$ Eq.(\ref{gammaedge2}) has only one solution $\gamma = 0$
corresponding to an unbound edge.  The transition
between the bound and unbound states actually happens before the limit
of stability is reached, namely when the energy (\ref{eedge2})
vanishes.  Numerical analysis shows that it happens at $\lambda \simeq
0.3412$.  This is close to the limit of stability thus implying that
the edge delocalization is a weak first-order transition.

The results of the analysis are summarized in Fig.~1 where we show the
bond stiffness $\gamma$ and the edge binding energy $E_{edge}$ as
functions of the quantum parameter $\lambda$.  The metastability
develops in the $1/4 \le \lambda \le 0.3730$ range:  for $\lambda <
0.3412$ the bound edge has lower energy while for $\lambda > 0.3412$
the ground state corresponds to a delocalized edge particle.              
\begin{figure}
\includegraphics[
  width=1.0\columnwidth,
  keepaspectratio]{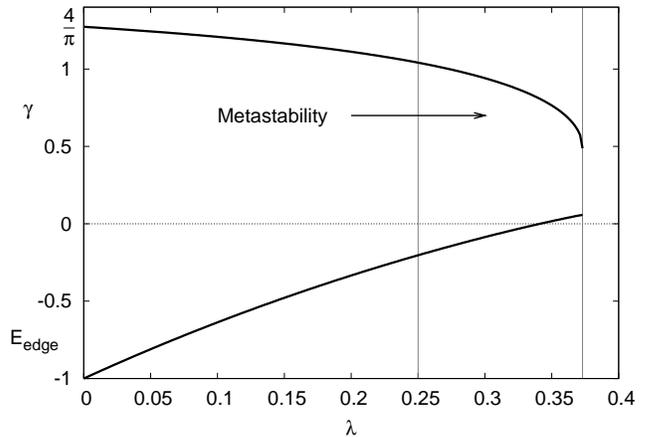}
\caption{Dimensionless stiffness of the edge bond $\gamma$ and
corresponding binding energy of the edge particle $E_{edge}$ of a
half-infinite Luttinger liquid as functions of the quantum parameter
$\lambda$ (\ref{lambda}).  The region of metastability
is confined to the $1/4 \le \lambda \le 0.3730$ range.}
\end{figure}

These conclusions should be contrasted with the properties of the
bulk Luttinger liquid.  Its range of existence is given by \cite{KQT} $0 <
\lambda \le 1.0591$ (or equivalently $0 < \lambda_{0} \le 0.3896$)
which is the condition that a solution $\lambda(\lambda_{0})$ to
Eq.(\ref{lambda_0versuslambda}) can be found for given De Boer's
number $\lambda_{0}$ (\ref{lambda_{0}}).  Therefore in the $0.3730 <
\lambda \le 1.0591$ range the bulk Luttinger liquid is stable against the
disordering effect of quantum fluctuations while the edge is not.
This is due to the stronger softening effect that zero-point motion has on the
free edge as compared to the bulk of the system.  The direct evidence
of this is presented in Fig.~2   
\begin{figure}
\includegraphics[
  width=1.0\columnwidth,
  keepaspectratio]{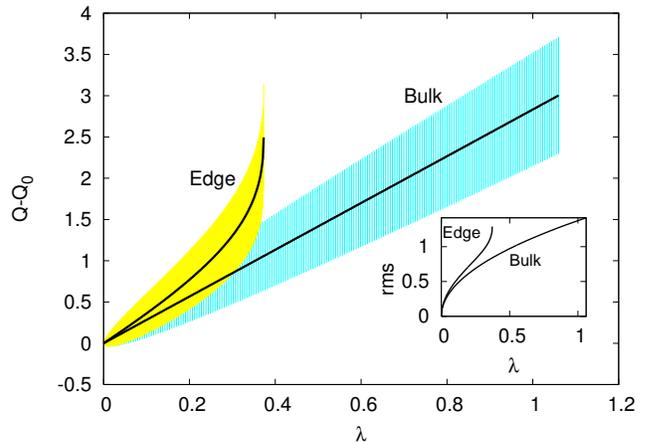}
\caption{(Color online)  Quantum expansion $Q - Q_{0}$ and rms
fluctuation of the bulk and edge bonds as functions of
the quantum parameter $\lambda$ (\ref{lambda}).  The rms fluctuation
is shown both as the vertical extent of shaded regions centered around
the quantum expansion curves, and explicitly in the inset.}  
\end{figure}
where we show the quantum expansion and rms fluctuation of the bulk
and edge bonds as functions of the quantum parameter $\lambda$
(\ref{lambda}) within their corresponding ranges of existence.  The
quantum expansion in the bulk (\ref{qexpansionbulk}) is a linear
function of $\lambda$ while the edge bond expands faster than linearly
because the bond stiffness $\gamma$ entering the argument of the
logarithm in (\ref{qexpansionedge}) is a decreasing function of
$\lambda$ as shown in Fig.~1.  Since the $\gamma(\lambda)$ dependence
is not very strong one can say that the edge expansion is roughly
twice the bulk value as suggested by the ratio of pre-logarithmic
factors in Eqs.(\ref{qexpansionedge}) and (\ref{qexpansionbulk}).
This can be understood by noticing that any bulk bond joins two
half-infinite Luttinger liquids thus implying that its dynamics is
twice as inertial as that of the edge.  In this sense zero-point
motion at the edge is about twice as strong as that in the bulk.  The
same argument explains why the edge rms fluctuation is roughly
square-root of two larger than its bulk counterpart (compare
Eqs.(\ref{rmsedge}) and (\ref{rmsbulk})).

In describing the dynamics of the edge bond the rest of the
system was approximated by a harmonic liquid with the bulk properties
which means that the bond adjacent to the edge has the length
and rms fluctuation identical to those in the
bulk.  This is an artifact and in reality, as one goes inside the
bulk, the bond lengths and their rms fluctuations decrease 
approaching the bulk values asymptotically.  This deficiency would be 
acceptable provided the calculated length of the edge bond and its rms 
fluctuation are not very different from their bulk counterparts.  
Since for molecular matter with pair interaction potential of the
Morse form the classical bond length satisfies the condition 
$Q_{0} \gtrsim 5$ \cite{KQT}, inspection of Fig.~2 shows that even
at the limit of its stability the length of the edge bond and its rms 
fluctuation do not exceed their bulk counterparts by more than
an acceptable $25\%$.  

Moreover, the relative fluctuation $<f^{2}>_{0~edge}^{1/2}/Q_{edge}$
is always significantly smaller than unity which implies that our
conclusions are weakly sensitive to the statistics of the underlying
particles and that the deficiencies of the Morse potential in
mimicking the true pair interaction at largest and shortest
distances are ignorable.  The latter allows us to argue that the edge
dissociation pre-emting the bulk instability is a general property of 
one-dimensional molecular matter.  
\begin{figure}
\includegraphics[
  width=1.0\columnwidth,
  keepaspectratio]{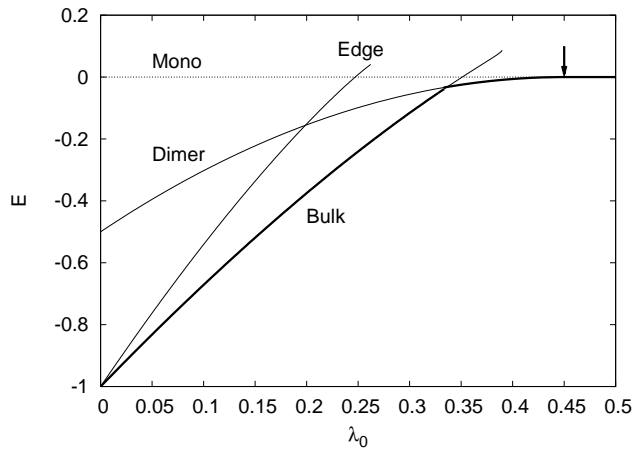}
\caption{The dependences of the energy per particle for various bulk phases of
the system on De Boer's number $\lambda_{0}$ (\ref{lambda_{0}})
together with edge binding energy.  The arrow pointing down is the
dimer dissociation threshold.} 
\end{figure}

In order to gain an insight into the consequences of this effect in Fig.~3
we plot  the ground-state energy per particle of the bulk
Luttinger liquid (given by Eqs.(\ref{lambda_0versuslambda}) and
(\ref{energybulk})) and the binding energy of the edge particle
(determined through Eqs.(\ref{lambda_0versuslambda}),
(\ref{gammaedge2}), and (\ref{eedge2}) as functions of De Boer's
number $\lambda_{0}$ (\ref{lambda_{0}}).  Additionally we show the
ground-state energy per particle for an infinitely diluted gas of Morse
dimers, 
$E_{dimer}(\lambda_{0})=
-(1/2)(1-\pi\lambda_{0}/\sqrt{2})^{2}$ \cite{Morse}.  The bold parts
of the curves describe the ground states of the bulk matter:
as De Boer's number increases, at $\lambda_{0} \simeq
0.3365 $ the Luttinger liquid evaporates via a discontinuous transition
into a gas of dimers followed by a continuous dissociation transition at
$\lambda_{0} = \sqrt{2}/\pi$ into a monoatomic gas \cite{KQT}.  For a
system with a free edge the binding energy of the edge 
particle $E_{edge}$ can become smaller than its dimer counterpart
$E_{dimer}$:  for $\lambda_{0} \gtrsim 0.1981$ the whole Luttinger liquid 
comes unraveled, \textit{two} particles at a time despite the fact
that the bulk condensed state is energetically favorable.  Since our
bulk and edge binding energies are variational upper bounds, in
actuality the dimer gas may not come into play; its role then will be
played by the monoatomic gas. 

If the escape of the edge particles to infinity is impossible due to
a distant obstacle, this will generate a vapor pressure and the bulk 
Luttinger liquid may coexist with a gas of particles.  As
$\lambda_{0}$ increases toward the point of the bulk
transition, dissociation proceeds inside the bulk in a manner similar
to that in surface melting \cite{SM}.  We hasten to mention
the speculative character of the statements of this paragraph which we 
plan to clarify in the future.   

The examples of one-dimensional matter with dissociated edge and
stable bulk, $0.1981 < \lambda_{0} < 0.3365$, include $H_{2}$ and
$D_{2}$ in free space, and more cases can be found in the presence
of a medium \cite{KQT}.   

This work was supported by the Chemical Sciences, Geosciences and Biosciences
Division, Office of Basic Energy Sciences, Office of Science, 
U. S. Department of Energy.

\end{document}